# Validity of crystal plasticity models near grain boundaries: a contribution of elastic strain measurements at the micron scale

Emeric Plancher[1], Pouya Tajdary[1], Thierry Auger[1,*], Olivier Castelnau[1], Véronique Favier[1], Dominique Loisnard[2], Jean-Baptiste Marijon[1], Claire Maurice[4], Vincent Michel[1], Odile Robach[3], Julien Stodolna[2]

[1] *PIMM/ENSAM-CNRS-CNAM, UMR 8006, 151 Bld de l'hôpital 75013 Paris, France.*
[2] *EDF, Département MMC, Avenue des renardières, 77818 Moret-sur-Loing, France*
[3] *CEA–CNRS CRG-IF BM32 beamline at ESRF, 6 rue Jules Horowitz, Cedex 9, 38043 Grenoble, France,*
[4] *Mines Saint-Etienne, Laboratoire Georges Friedel, UMR CNRS 5307, 158 cours Fauriel, 42100 Saint-Etienne, France*

**Abstract**:
Synchrotron Laue microdiffraction and Digital Image Correlation measurements were coupled to track the elastic strain field (or stress field) and the total strain field near a general grain boundary in a bent bicrystal. A 316L stainless steel bicrystal was deformed *in situ* into the elasto-plastic regime with a four-point bending setup. The test was then simulated using finite elements with a crystal plasticity model comprising internal variables (dislocation densities on discrete slip systems). The predictions of the model have been compared with both the total strain field and the elastic strain field obtained experimentally. While activated slip systems and total strains are reasonably well predicted, elastic strains appear overestimated next to the grain boundary. This suggests that conventional crystal plasticity models need improvement to correctly model stresses at grain boundaries.

## 1. Introduction

The buildup of intergranular microstresses (type-II internal stresses) and intragranular stresses or residual stresses (type-III internal stresses) upon plastic deformation is a central issue in mechanics of materials [1]. The development of local stresses is generally due to incompatibilities between the anisotropic (elastic and plastic) deformation on both sides of grain boundaries. Another contribution to internal stresses is associated with spatial localization patterns at the microstructural scale of plastic deformation. Since plasticity precedes many other metallurgical phenomena such as recrystallization, phase transformations or damage, its correct understanding (*i.e.* the ability to reliably predict local stress/strain fields in a plastically deformed material) has major implications. For example, the simulation of intergranular stresses is critical to describe environmental effects such as stress corrosion cracking or liquid metal embrittlement, because intergranular stresses drive the dynamics of fracture [2, 3, 4]. During recrystallization, the stored elastic energy associated with the dislocation network governs the microstructure evolution [5, 6]. It is therefore desirable to quantify the degree of accuracy on local stress/strain fields one can reach through simulation, by confronting experimentally accessible stress and strain fields to modelling results in polycrystals.

The usual measurement of internal stresses is performed with X-ray or neutrons diffraction, by using beam cross section significantly larger than the grain size. Those techniques often lead to a reasonably accurate but averaged estimation of micro-strains [7] that compares well with plastic texture simulations [8,9,10]. Recently, the advent of local probes such as digital image correlation (DIC) [11], high angular resolution electron backscatter diffraction (HR-EBSD) [12,13] or synchrotron Laue microdiffraction [14,15] has opened the way to a new class of measurements that can directly map mechanical fields to a high degree of accuracy, down to a micrometer spatial resolution. Therefore, full field measurements can be carried out at the same scale than microstructural models featuring crystal plasticity constitutive laws, and paves the way to advanced validation of scale-transition modelling.

---

* Corresponding author, thierry.auger@ensam.eu

One of the key questions raised when using crystal plasticity models is whether grain boundaries can be treated as sole geometrical interphases, as thin surfaces with specific properties and special kinematics conditions [16], or as non-local microstructural constituents with special properties *per se* [17]. Grain boundaries are obstacles to dislocations slip but, depending upon their nature or upon the misorientation between adjacent grains, they may still transmit some amount of plastic deformation from one grain to the other [18]. Conventional crystal plasticity models, such as the one employed in this publication [19, 20], do not define any particular behavior for the grain boundary, an assumption that may not be realistic enough.

In this work, a concentrated austenitic solid solution, typical of 316L stainless steel, is employed as a model alloy. We report here an original comparison between two mechanical fields measured at similar and fine spatial resolution, *i.e.* elastic and total (elastic + plastic) strain fields versus the prediction of a local crystal plasticity simulation [19]. First, the crystal plasticity parameters were fitted for the alloy and the set of parameters was validated by considering experimental fields developing in a single crystal [21, 22]. Then, an *in situ* bending experiment was performed at the European ESRF synchrotron to investigate how both strain fields localize near a general grain boundary in a large bi-crystal. Strain measurements were performed near the grain boundary at several loading steps with optical DIC (total strains) and Laue microdiffraction (elastic strains). The agreement between the prediction of the finite element simulation and the experimental data is discussed for both strain fields and the local activation of slip systems.

## **2. Experimentals**
### *2.1 Materials*

Macroscopic single crystals of high purity austenitic stainless steel were grown by Bridgman directional solidification in a horizontal furnace with an argon cover gas. For tensile and bending experiments, single crystal samples were cut by spark erosion from a macroscopic single crystal, with the crystal axes aligned with the sample axes within 3°. A bi-crystal bending sample was also cut out from a macroscopic oligocrystal obtained in a separate crystal growth attempt. The grain boundary was selected for its position, considering geometrical requirements for the bending sample, and thus its character was random. Therefore, the chemical composition of the bicrystal and the single crystals slightly differs (see table 1). The gauge section of the two tensile specimens was 8x3x0.5 mm$^3$. The shape of the single crystal bending specimen was 30x4.3x0.46mm$^3$. The shape of the bicrystal bending specimen was 30x4.8x0.48mm$^3$. All samples were carefully polished to a 1µm grade or below and the preparation quality checked using channeling contrast imaging in a scanning electron microscope. Details of the single crystal bending and tensile experiments have been reported in [22]. The elastic constants of the 316L oligocrystal were measured by Resonant Ultrasound Spectroscopy and reads: $C_{11}$=202 GPa, $C_{12}$=130 GPa and $C_{44}$=128 GPa (uncertainty on the values is evaluated below +-1GPa).

| Wt% | Fe | Cr | Ni | Mo | C | Mn, N, S, P, Si, Cu, O |
|---|---|---|---|---|---|---|
| Monocrystal tensile | Bal. | 17 | 14.6 | 2.3 | <0.002 | Not detected |
| Monocrystal 4-pts bending | Bal. | 17 | 14.6 | 2.3 | <0.002 | Not detected |
| Bi-crystal 4-pts bending | Bal. | 15.5 | 14 | 1.7 | <0.002 | Not detected |

Table 1: Chemical composition of the high purity alloys (single crystal and bi-crystal).

*2.2 Laue microdiffraction and DIC under synchrotron radiation*

*In situ* Laue microdiffraction experiments were carried out at the French CRG beamline BM32 of the European Synchrotron (ESRF, Grenoble, France), described in reference [23]. The incoming polychromatic X-ray beam (energy range 5-23 keV) was focused down to ~800x600 nm$^2$ by means of Kirkpatrick–Baez mirrors. Laue pattern were recorded using a MARCCD® detector (pixel size: 79.14 microns, 2048x2048 pixels, 16 bits dynamic) positioned 60mm above the specimen surface. A strain-free Ge wafer was used to calibrate the setup geometry (detector orientation, etc.). A custom-designed 4-point bending rig was installed on the translation stage pre-tilted at 40°. The device was equipped with a load cell with a capacity of 20N and an extensometer sensitive to displacements below 1µm to measure the macroscopic force/displacement imposed on the sample.

An optical microscope available on beamline BM32 (Allied Vision Technology GiGE®, Objective Mitutoyo 10x/0.28, f=200) was employed *in situ* to acquire optical images used for the measurement of the total strain field on the specimen cross-section surface by the DIC technique. A MoS$_2$ powder (MOLYKOTE® Microsize, with submicron grain size) was deposited on the sample surface to create the random speckle pattern. When a DIC image is to be acquired, the optical microscope has to be positioned such that its focal plane covers the sample cross-section surface. Therefore, measurements were carried out sequentially, the microscope being brought in and out between loading steps. Further details of the experimental setup can be found in [22].

Processing of the Laue microdiffraction data was performed with the software LaueTools (see e.g. [24]) to extract the deviatoric part of the elastic deformation tensor [25]. In this work, the full elastic strain tensor was recovered assuming a vanishing stress vector on the specimen surface, which is a sound approximation due to the small attenuation length of X-rays (65 microns in Fe at 22 keV) and the lack of in-depth microstructure gradients in the specimen. The plane stress hypothesis was used to determine the hydrostatic part of the stress tensor and then, using generalized Hook's law, the hydrostatic part of the elastic strain tensor was extracted. Laue patterns consisted mostly in a single diffraction figure, except on the grain boundary where two overlapping figures could be detected. In that case, the figure with the more intense diffraction peak was indexed and associated to the relevant grain. The area investigated on the sample was approximately 400µmx600µm, centered on the grain boundary with a step size of 25µm. Orientation angles use the Bunge convention in degrees throughout the paper ($\varphi_1, \Phi, \varphi_2$). The maximum measured lattice rotation between the initial state and the loaded state reaches 2.5° point to point (in the scanned area).

## 3. Modelling

*3.1 Formulation*

The kinematics of the crystal plasticity model relies on the finite transformation framework (small elastic distortions but large lattice rotations) first proposed by Lee where the deformation gradient tensor is built as the product of an elastic part with a plastic part [26]. The elastic part embodies the anisotropic elastic stretch and the grain rotation while the plastic one accounts for crystallographic slipping along specific slip systems (up to 12 for a fcc crystal). The constitutive laws rely on dislocation densities resolved on each slip system as internal variables $\rho^{su}$ whose evolution model strain hardening in metals and alloys. At each time increment of the computation, the local critical shear stress $\tau_c^s$ is computed with the updated dislocation densities using the forest hardening interaction matrix resolved on every slip system (via the $a^{su}$ interaction matrix between slip systems (s) and (u) including diagonal self-interaction terms) [27]:

$$\tau_c^s = \tau_0 + \mu b \sqrt{\sum_{u=1,12} a^{su} \rho^{su}} \qquad (1)$$

where $\mu$ is the isotropic shear modulus, *b* the norm of the Burgers vector and $\tau_0$ the lattice friction stress. The onset of slip is triggered according to the Schmid law. The rate-dependent slip rate prescription proposed by Pierce [28] for single crystals is used. It is approximated using a power law:

$$\dot{\gamma}^s = \dot{\gamma}_0 \left(\frac{|\tau^s|}{\tau_c^s}\right)^n sign(\tau^s) \; if \; |\tau^s| > \tau_c^s, \; \dot{\gamma}^s = 0 \; otherwise \qquad (2)$$

where $\dot{\gamma}_0$ is a reference shear rate and $n$ is the stress sensitivity parameter. These material parameters are to be fitted to reproduce the correct flow stress rate dependence. Following Teodosiu's law with the hypothesis of a constant mobile dislocation density [19], the rate of evolution of the stored dislocation density is governed by a dislocation production term and is balanced by a dislocation annihilation term taking into account the dynamic recovery during deformation (related to the annihilation distance of dislocation dipoles $y_c$):

$$\dot{\rho}^s = \frac{|\dot{\gamma}^s|}{b}\left[\frac{\sqrt{\sum_{u \neq s}\rho^u}}{K} - 2y_c\rho^s\right] \qquad (3)$$

where $K$ and $y_c$ are material's parameters that are usually fitted to reproduce stage I and stage II on a single crystal. The approach is implemented in the Abaqus® finite element code, using a UMAT subroutine [29].

*3.2 Parameter identification*

Crystal plasticity parameters for 316L stainless steels can be found in several works to date [4,30]. There are three categories of parameters. The first one comprises physical parameters that can be set based on the metallurgical state of the sample (burgers vector, initial dislocation density $\rho_0$ and initial critical friction stress $\tau_0$). (As said previously the 316L elastic constants are taken as: $C_{11}$=202 GPa, $C_{12}$=130 GPa, $C_{44}$=128 GPa). The second set of parameters allows to describe the viscous or the rate dependent effects ($\dot{\gamma}_0$ and $n$). For 316L stainless steel, it is important to set carefully these parameters because austenitic steels are known to have a non-vanishing rate-dependent flow stress that becomes negligible only below strain rate of $10^{-2}$ s$^{-1}$ [31]. For our purpose of testing at slow strain rate (i.e. lower than $10^{-2}$ s$^{-1}$), these parameters were constrained to correctly reproduce a quasi-static loading state with a minimized strain rate effect ($\dot{\gamma}_0 > 10^{-5} \; s^{-1}$ and $n > 20$). In order to perform this identification, tensile tests carried out on single crystals at a strain rate of $10^{-4}$ s$^{-1}$ along direction [100], as detailed in [22], were modelled by finite elements in Abaqus using C3D8 elements with a [100] crystal orientation. The parameters $\tau_0$, $\dot{\gamma}_0$ and $n$ were then adjusted to fit the experimental curves (Figure 1). The smooth transition at the onset of plasticity given by the Pierce strain rate law is damped due to our choice of parameters.

Figure 1: Elasto-plastic response of a single crystal deformed under uniaxial tension along direction [100]. Experimental data is compared with the identified constitutive relation.

The third category of parameters (describing strain hardening) is composed first of the $a^{su}$ interaction matrix between the slip systems and the two material parameters $K$ and $y_c$. The interaction matrix ($a_0$ for self-interaction, $a_1$ for collinear interactions, $a_2$ for Lomer Cottrell locks, $a_3$ for Hirth junctions, $a_4$ for glissile dislocation interactions, $a_5$=$a_0$ for sessile dislocation interactions) is taken from the work of Devincre [32] which was later on already applied for the 316L(N) stainless steel [30]. The two remaining parameters were set to reproduce the slope of the single crystal tensile curves reported in Figure 1. It is known however that K and $y_c$ may vary with the strain level and our choice may need to be refitted to better reproduce a larger amount of experimental data.

Overall, the resulting parameters are charted in table 2 along, when applicable, with their values as found in two other works in which a rate-dependent crystal plasticity approach was used [4, 30]. The major differences between different parametrizations can be rationalized by being due either to a compositional difference with our alloys or to a different initial dislocation density ($\rho_0$ and $\tau_0$ play in synergy to give a larger flow stress than our well annealed alloys).

| | b (m) | $\rho_0$ (m$^{-2}$) | $\tau_0$ (MPa) | $\dot{\gamma}_0$ (s$^{-1}$) | $n$ | K | $y_c$ (m) | $a_0$ | $a_1$ | $a_2$ | $a_3$ |
|---|---|---|---|---|---|---|---|---|---|---|---|

| | | | | | | | | | | |
|---|---|---|---|---|---|---|---|---|---|---|
| Our work | 2.54x10$^{-10}$ | 1.6x10$^{10}$ | 22.3 | 10$^{-4}$ | 40 | 18 | 1.25x10$^{-9}$ | 0.1388 | 0.633 | 0.079 | 0.1236 |
| [30] | 2.54x10$^{-10}$ | 1.72x10$^{12}$ | 22.3 | 4.10$^{-11}$ | 73.5 | 59.97 | 1.29x10$^{-9}$ | 0.1388 | 0.633 | 0.079 | 0.1236 |
| [4] | | N.R. | 90 | 10$^{-3}$ | 55 | N.A. | N.A. | N.A. | N.A. | N.A.* | N.A.* |

**Table 2:** Crystal plasticity physical parameters used for the 316L stainless steel (N.A. Not Applicable, N.R. Not Reported, when the unit is not specified, the parameter is adimensional. * We note that there is an inversion between a$_2$ and a$_3$ in [30] compared to [32])).

This parametrization was validated against the measurement of the elastic strain and total strain fields, performed by Laue microdiffraction and DIC respectively along the specimen cross-section, in the bending experiment of a 316L single crystal (see [22]). For the simulation, the dimensions of the sample were 0.46×4.3×30 $mm^3$ and the [100] orientation was aligned with the loading axis. The four point bending test configuration was simulated by adding four rigid cylindrical shells (outer shells are fixed; inner shells move along the *Ox* axis). The contact interaction conditions between the four pins and the single crystal beam were frictionless sliding. The imposed boundary condition displacement rate used for the simulation was set to be in the quasi-static strain rate domain of the crystal plasticity model. The beam deflection was simulated up to the final measured one. One can see in figure 2 that the total strain and elastic strain profiles were reproduced rather accurately along the cross-section of the sample. The stress-strain curve of the single crystal oriented for multiple slips did not present stage I, but stage II. It reproduces, nicely the transition from elastic to plastic deformation together with the slope of elastic strain in the plastic regions associated with the crystal hardening. This gives us confidence that the crystal plasticity parameters are correctly set according to the fitting strategy that was used in this work.

Figure 2: Elastic and total (i.e. elastic + plastic) strain evolution across the thickness of a [100] bent single crystal. The position of measurement points are along direction *x* on the surface normal to *z* (see inset). Experimental data from DIC and Laue microdiffraction (grey curves) are compared to the identified crystal plasticity model (blue and red curves).

**4. Experimental and modelling setup for the bicrystal**

A schematic view of the investigated bicrystal can be found in figure 3. The supposedly planar grain boundary was inclined with respect to the specimen axes as indicated. The misorientation angle between the two grains is 64° specifying a general grain boundary. The specimen was loaded *in situ* in the four-point loading device using the setup described in section 2.2. Sample deformation was carried out step by step and measurements were performed at 0 N (initial state), 3 N, 7 N, and then after complete unloading. After each load increment, the pin displacement was blocked, and we waited until the specimen relaxation was complete (~20 min). Then, a Laue map covering the area of the grain boundary was carried out. A DIC mapping measurement was also acquired at each Laue step while keeping the sample under load. The lower and the upper grains are respectively labelled Grain 1 and Grain 2.

Figure 3: Schematic of the four-point bending experiment performed on a 316L bicrystal and details of the FEM mesh next to the grain boundary. Euler angles for grains G1 and G2 are (-175°, 78°, -107°) and (-133°, 25°, -163°) respectively.

The modelling by crystal plasticity of the bicrystal bend test was performed similarly to the single crystal by simulating the entire bicrystal in Abaqus (dimensions 0.48x4.8x30 $mm^3$). The beam was split in two parts with a slanted surface representing the grain boundary (at an angle of 50° in the *xy* plane and 37° in the *yz* plane). The mesh for the finite element modelling was then generated using C3D8 elements with a fine grid close to the grain boundary and a coarser grid further away to save computational power. The fine/coarse separation was set manually to model with a coarse mesh the part of the beam that stays in the elastic regime so that the entire plastically deformed volume was covered with a fine mesh. The four point bending test configuration is simulated by adding four rigid cylindrical shells with identical boundary conditions and contact interactions. The displacement rate used for the simulation was small enough to be in the quasi-static loading state of the crystal plasticity model. The total strain was given as a result of the non-linear finite element modelling while the elastic strain was extracted at each step by elastically unloading the current stress using the generalized Hooke's law in the crystal reference frame. The elastic strain tensor was then rotated back to the sample reference frame to be readily compared with the measured one.

### 5. Results on the bicrystal:

After complete unloading, the pattern formed by the MoS$_2$ deposit used for DIC at the surface of the bicrystal was cleaned in an ultrasonic bath. The sample surface was then observed using channeling contrast in a scanning electron microscope adapted for large field of view observations with low image distortions (TESCAN MIRA3®). The backscattered electron image reported in figure 4 reveals the grain boundary and multiple slip traces observed in both grains. In grain 2, a slip trace analysis using the orientation data from Laue microdiffraction ($\varphi_1 = -133°$, $\Phi = 25°$, $\varphi_2 = -163°$) can provide the slip planes associated with the four families of traces observed. In grain 1 ($\varphi_1 = -175°$, $\Phi = 78°$, $\varphi_2 = -107°$) the slip traces associated with slip planes (-1-11) and (1-11) are very close with a theoretical misorientation in the surface plane of only a few degrees.

To compare with the slip systems activated in the simulation, the area of interest was schematically sliced into 12 domains and the number of active systems tallied. The division was based on (i) approximate distance from the grain boundary and (ii) the nature of the applied stress (tensile, compressive or near neutral fiber). The activated slip systems and its corresponding dislocation density were extracted from the simulation as state variables. In the domains were a slip system is not activated ($|\tau^s| < \tau_c^s$ or $\dot{\gamma} < 10^{-6} \, s^{-1}$ at all steps), the corresponding dislocation density variable remains constant. A visual one to one comparison between the slip traces in each domain from figure 4a and the activated slip systems given by a Schmid factor analysis was performed. It shows that in all the regions except 6 and 9 (which are closer to the grain boundary and are in the area which is subject to tensile stress) the simulation predicts correctly the type and number of activated slip systems. In the tensile part of the sample, according to the simulation, Grain 1 should be in dual slip mode (111)<01$\bar{1}$> and (1$\bar{1}$1)<011> while grain 2 should be closer to single slip mode ($\bar{1}\bar{1}$1)<101>. However in the experiment, due to internal stresses, not every predicted slip plane is activated. Grain 1 is rather deformed and grain 2 tries to accommodate.

Figure 4: Backscatter electron image of the area of interest after complete unloading. **(a)** Slip traces patterns are detected in both grains revealing single or multiple slip activation. A slip trace analysis has been performed to identify the slip planes associated with the different families of slip traces. **(b)** Division of the area into 12 regions used to compare with the finite element model.

The experimental strain fields (total and elastic) have been extracted following the procedure described earlier respectively from the DIC and Laue diffraction data. They are plotted for the *yy* component (along the *y* axis of figure 3) of the total strain tensor (figure 5) and of the elastic strain tensor (figure 6). The total strain $\epsilon_{yy}^t$ distribution shows that the right part of the specimen is mainly in tension and the left part is mainly in compression as expected from bending. However, the grain boundary clearly disturbs the strain field. The strain magnitude is stronger in Grain 1 than in Grain 2. The simulation results for $\epsilon_{yy}^t$ are represented in the figure 5b. One can see that the main plastic activity domains are very well reproduced except at the top right corner. The normalized difference plot is shown in figure 5c. It reveals that the difference between the measurement and the simulation results over almost the entire analyzed surface, including next to the grain boundary area, remains below 20%. This level of agreement is considered as being rather good.

Figure 5: Total strain fields along *y* axis: experimental, modelling and difference plot at a loading of 7 N

The elastic strains $\epsilon_{yy}^e$ are shown in figure 6. One can see that the experimental elastic strains follow closely what is expected from a bending beam experiment. One sees clearly an area in tension at the right and an area in compression at the left consistent with the existence of a neutral fiber in a four-point bending experiment. The effect of the grain boundary can be seen weakly as some higher *yy* strain component appear as hot spots along the grain boundary. In figure 6b one can see the simulated corresponding elastic strain. Overall the shape looks similar but the intensity scale is significantly different and appears globally to be of higher magnitude than the experimentally measured one. The normalized difference plot is shown in figure 6c. The presence of a grain boundary can be easily inferred from this plot with two main sources of disagreement. The first one is the presence of an area at the right of Grain 1 (the area in tension) that seems to be frequently in disagreement between experimental data and simulation results. The other stunning characteristics is the clear distinction of the grain boundary as a major source of contribution of disagreement between experimental data and simulation results. This indicates that next to the grain boundary is a special area where crystal plasticity somehow overestimates the elastic strains induced by the plastic activity due to bending. The elastic strain from a large part of Grain 2 is well predicted, especially away from the grain boundary. This is expected as the area was only slightly plastically deformed (<0.5%). The same conclusion can be reached for Grain 1 close to the neutral fiber, but to a lesser extent.

Figure 6: Elastic strain fields along *y* axis: experimental, modelling and difference plot at a loading of 7 N

## 6. Discussion

The *in situ* mapping of total and elastic strains by DIC and Laue microdiffraction allowed to characterize the evolution of these fields around a general grain boundary, with a model alloy of 316L, in a beam bending configuration. This constitutes a major step forward compared to similar studies aiming at the measurement of elastic strains, such as those carried out a few years ago which had a beamsize of 100 μm x 100 μm compared with 1 μm x 1 μm here [33] or what can be achieved with high energy X-ray diffraction [34]. The currently achieved spatial resolution with Laue microdiffraction at synchrotron facilities allows to tackle this issue with much higher accuracy.

The results of this work show that the heterogeneities in active slip planes are very well predicted with the formation of well identified plastic domains in the different parts of the area surrounding the grain boundary. While some localized discrepancies arise at one edge of the specimen (more specifically at the right-hand side in the tensile area), one can note that the difference between the predicted domains and the observed ones are not specifically related to the grain boundary. The total strain field is also very well predicted. It is mainly controlled by the macroscopic kinematics but the presence of grain boundary and the induced change in anisotropy clearly disturbs the field around it. This work brings another example to the well-known effect of grain boundaries on the formation of plastic heterogeneities or plastic glide domains [35,36]. It also illustrates the ability of standard crystal plasticity modelling to predict in a detailed manner these early stages of plastic deformation (the maximum total strain is here less than 2%). The measured elastic strain field has mainly the symmetry induced by the bending geometry but modified, to a lesser extent, by the presence of the grain boundary. Its prediction by conventional crystal plasticity is, on the other hand, much less accurate than for the total strain. Elastic strains and so internal stresses are sensibly overestimated in the neighborhood of the grain boundary and in the right-hand corner of the analyzed area (grain 1). Away from the grain boundary it is predicted correctly with an accuracy below 20% over more than 50% of the analyzed area.

Let us first discuss possible experimental artifacts. A possible source of discrepancy between the crystal plasticity model and experimental data could come from the relaxation of the specimen. Here, elastic and total strains were measured when relaxation was terminated and the applied force was stabilized. Typical stress drop between the end of the loading step and the relaxed state lies in the 5-20% range (see [22]). Relaxation is associated with limited glide of a small number of dislocations to adjust and minimize internal stresses (micro-plastic regime). It has therefore a limited influence on the total strain field but affects the elastic one in a significant manner. Relaxation is a standard concern for *in situ* stress measurements, observed on many specimens such as bulk polycrystals [37] or thin films [38] but also in 316L austenitic steel at room temperature [39]. In austenitic steels, it has been reported that the early stages of plastic deformation involve first the activation of lattice dislocations then followed by the activation of triple point sources (absent in our case) and the formation of dislocation pile-ups thereafter slip transmission across the grain boundaries [40]. Scanning electron microscope imaging at high resolution of the grain boundary area showed typical direct slip bands transfer across the grain boundary which would support this interpretation and would explain why relaxation occurs more specifically near the grain boundary. Recently, a polycrystal mean-field homogenization model coupled with a dislocation-based constitutive relation accounting for statistical distributions of internal stresses has also been proposed to reproduce the effect of relaxation (polycrystal average) on neutrons diffraction data [41]. This approach was however not concerned with the spatial resolution as the one used here with Laue experiment and full-field finite element modelling. A relaxation effect linked to grain boundaries was not envisioned and it relies on an additional parameter, the standard deviation of the distribution of resolved shear stress. On the experimental side, further investigation of this issue is required that should now be possible when using the new X-ray detector of BM32 operating at 10Hz. One could try to monitor the time evolution of elastic strains near a grain boundary for example. It might also be required to study at the micron scale from the interface the slip transfer such as carried out in several recent works [42, 43].

Only the results concerning the *yy* component of the stress and strain fields are presented. Laue data were analysed using the standard method implemented in LaueTools software, in which the position of Laue spots on the X-ray detector screen were compared with the theoretical ones computed for an elastically strained Fe crystal lattice. As shown in [44, 45, 46], the digital image correlation technique can be used to estimate very precisely (accuracy of few hundredths of detector pixel) the spot displacement between two loading steps, leading to enhanced accuracy in the stress estimation. On ideal crystals (Si and Ge single crystals), stress accuracy of the order of 1MPa could be reach with this Laue-DIC technique. This constitutes another possible continuation of this work, allowing the investigation of the grain boundary effects on other components of the stress field. This will make it also possible to study lattice rotations with an accuracy better than $10^{-4}$ rad.

The results presented here suggest that grain boundaries as microstructural objects need a special treatment in crystal plasticity modelling such as an accurate definition of slip transfer following [47], eventually complemented by the effect of incompatibility strains [48, 42] and/or strain gradient effects [49]. Applying jump conditions for geometrically necessary dislocations may also be required [50]. Strain rate effects on the slip transfer could also be considered [51]. It could be required to treat the grain boundary as a thin surface with specific properties or special kinematics conditions allowing sliding and torsion [16]. Another class of interfaces treatment could be to integrate grain boundaries as a non-local microstructural constituent with special properties *per se* [17]. Going further would require to compare the results on our geometry with several of these approaches.

**7. Conclusion**

In trying to validate elastic strains (e.g. internal stresses) simulated by crystal plasticity with measured ones, it is shown that they are sensibly overestimated in the neighborhood of a general grain boundary, while there are better reproduced away from the grain boundary. A refinement of such type of experiment would allow to exclude artifacts or guide possible extensions of crystal plasticity models (inclusion of relaxation or slip transfer at grain boundaries). One could also explore the extension of modelling to various more sophisticated crystal plasticity models in order to be able to fully reproduce at least the bicrystal case. On the application side, this work demonstrated that the question of internal stresses due to incompatibility of the deformation at the grain boundary is not yet settled, hindering a thorough understanding of intergranular stress-controlled effects that remains a challenge in materials science.

**Acknowledgement**

C. Rey is warmly thanked by the authors for many fruitful discussions and for allowing us to use the CristalECP code. The "Laboratoire d'Imagerie Biomédical" from Sorbonne Université (Paris, France) is acknowledged for its help in measuring elastic constants by ultrasonic means. Beamtime allocation at the French Beamline BM32 CRG-IF at the European Synchrotron Facility (ESRF) is gratefully acknowledged (award No. HC/913 & HC/1449).

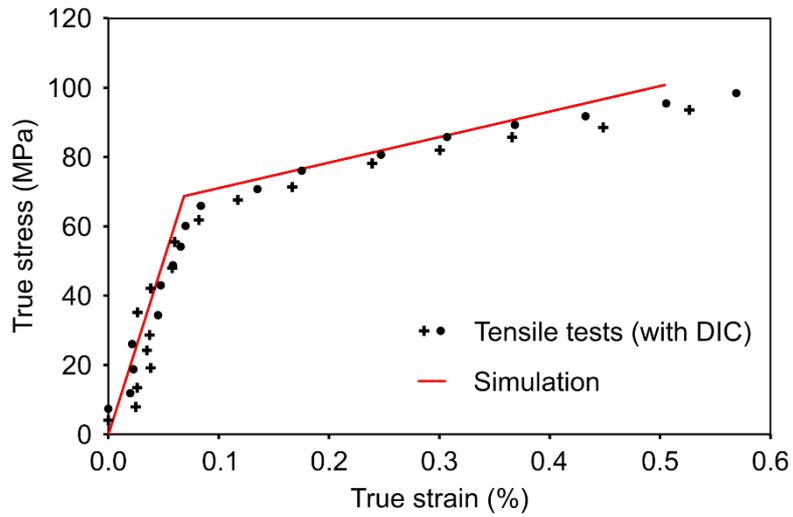

Figure 1: Elasto-plastic response of a single crystal deformed under uniaxial tension along direction [100]. Experimental data is compared with the identified constitutive relation.

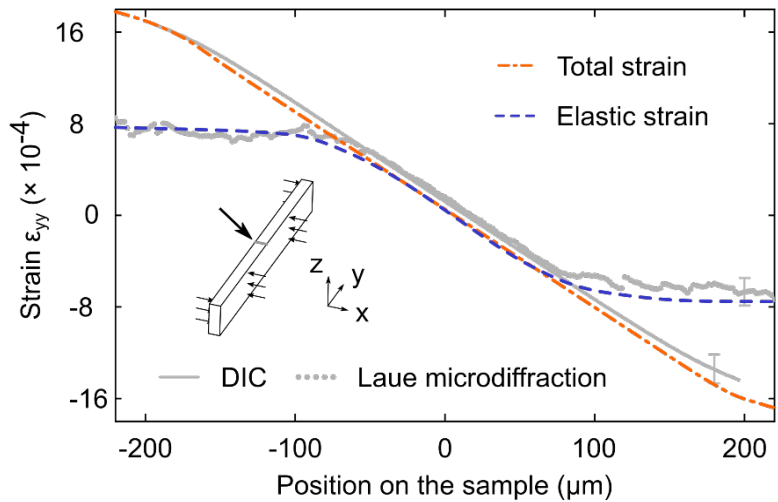

Figure 2: Elastic and total (i.e. elastic + plastic) strain evolution across the thickness of a [100] bent single crystal. The position of measurement points are along direction $x$ on the surface normal to $z$ (see inset). Experimental data from DIC and Laue microdiffraction (grey curves) are compared to the identified crystal plasticity model.

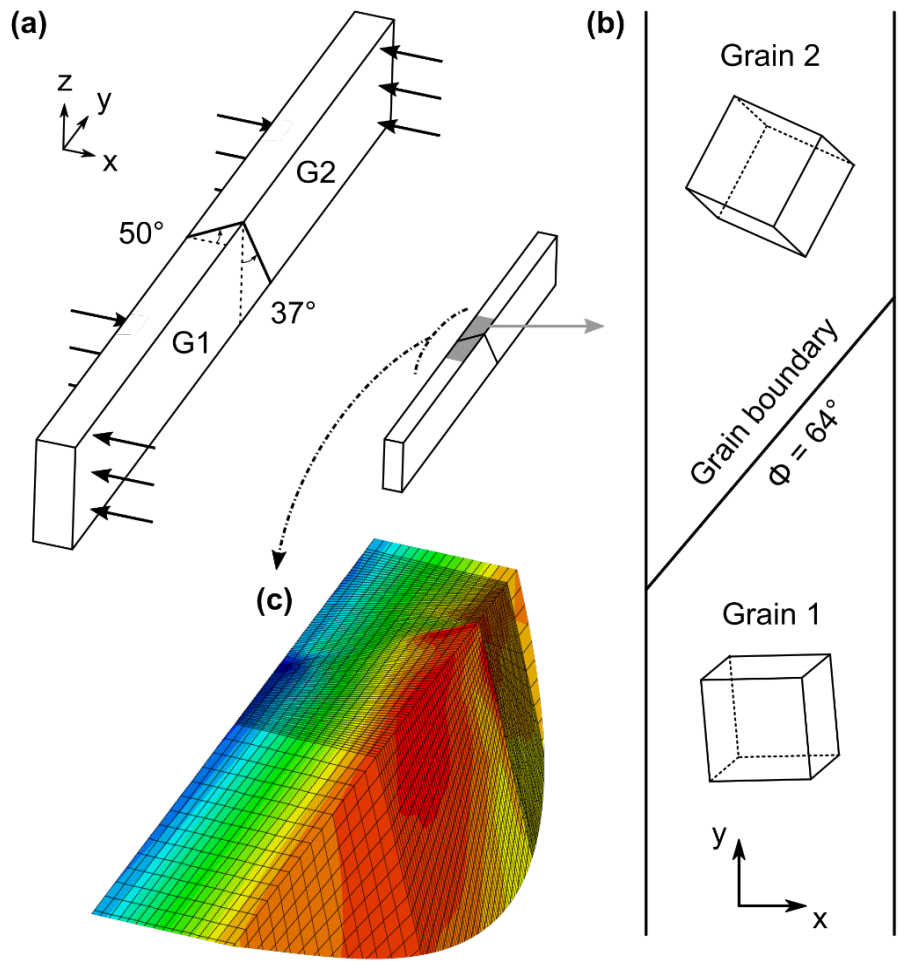

Figure 3: Schematic of the four-point bending experiment performed on a 316L bicrystal and details of the FEM mesh next to the grain boundary. Euler angles for grains G1 and G2 are (-175°, 78°, -107°) and (-133°, 25°, -163°) respectively.

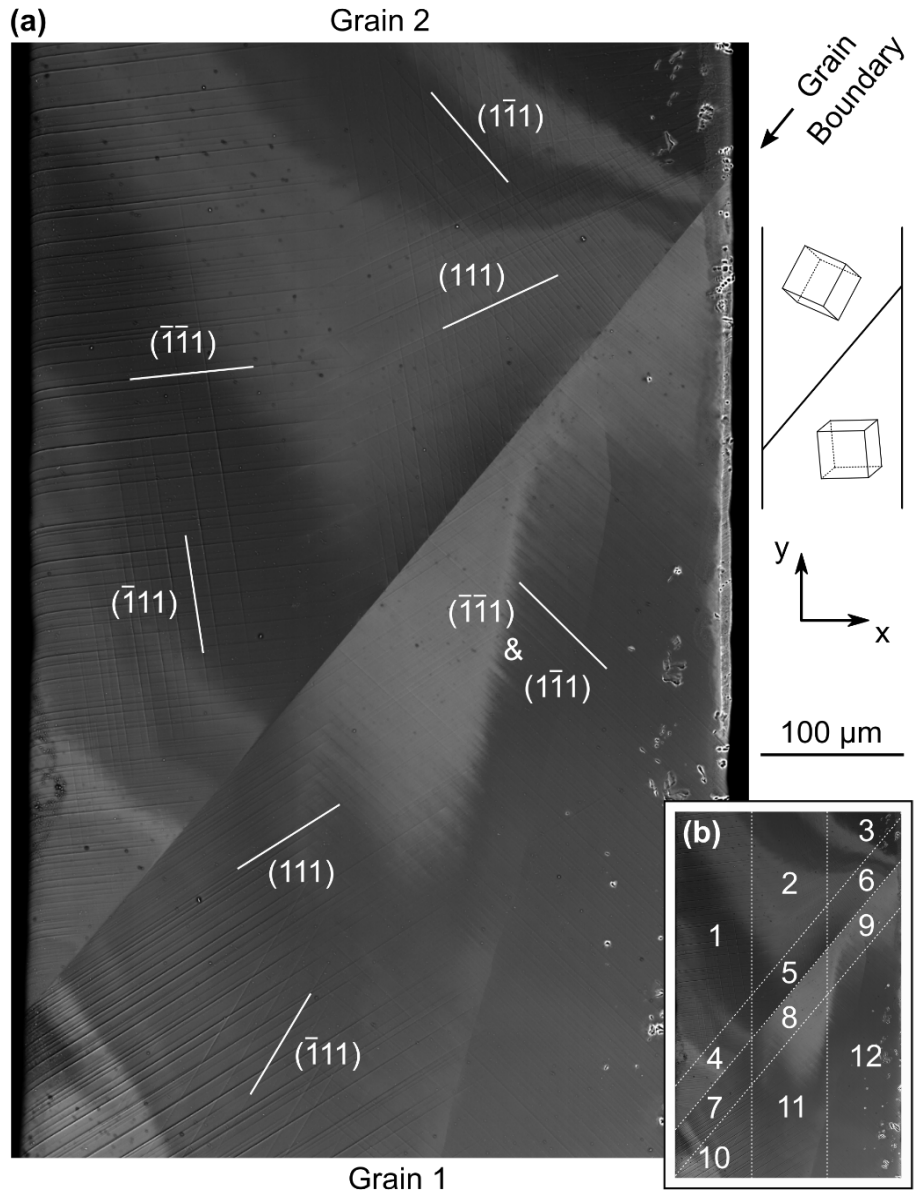

Figure 4: Backscatter electron image of the area of interest after complete unloading. (a) Slip traces patterns are detected in both grains revealing single or multiple slip activation. A slip trace analysis has been performed to identify the slip planes associated with the different families of slip traces. (b) Division of the area into 12 regions used to compare with the finite element model.

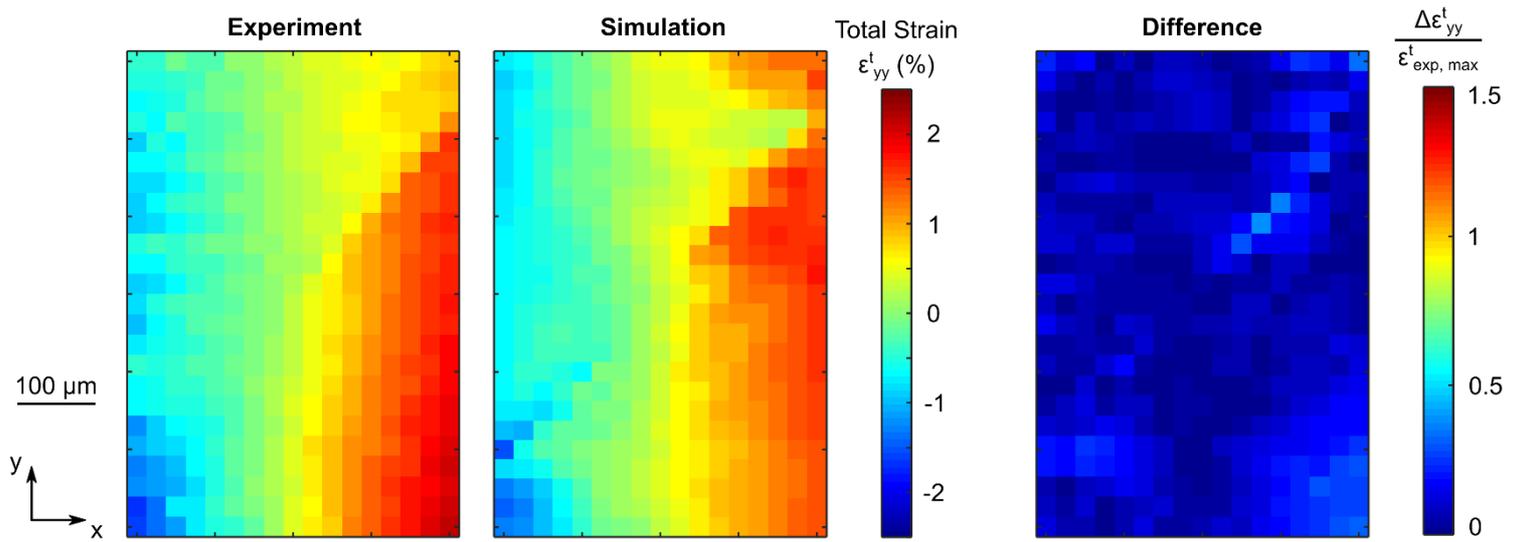

Figure 5: Total strain fields along the y axis: experimental, modelling and difference plot at a loading of 7N

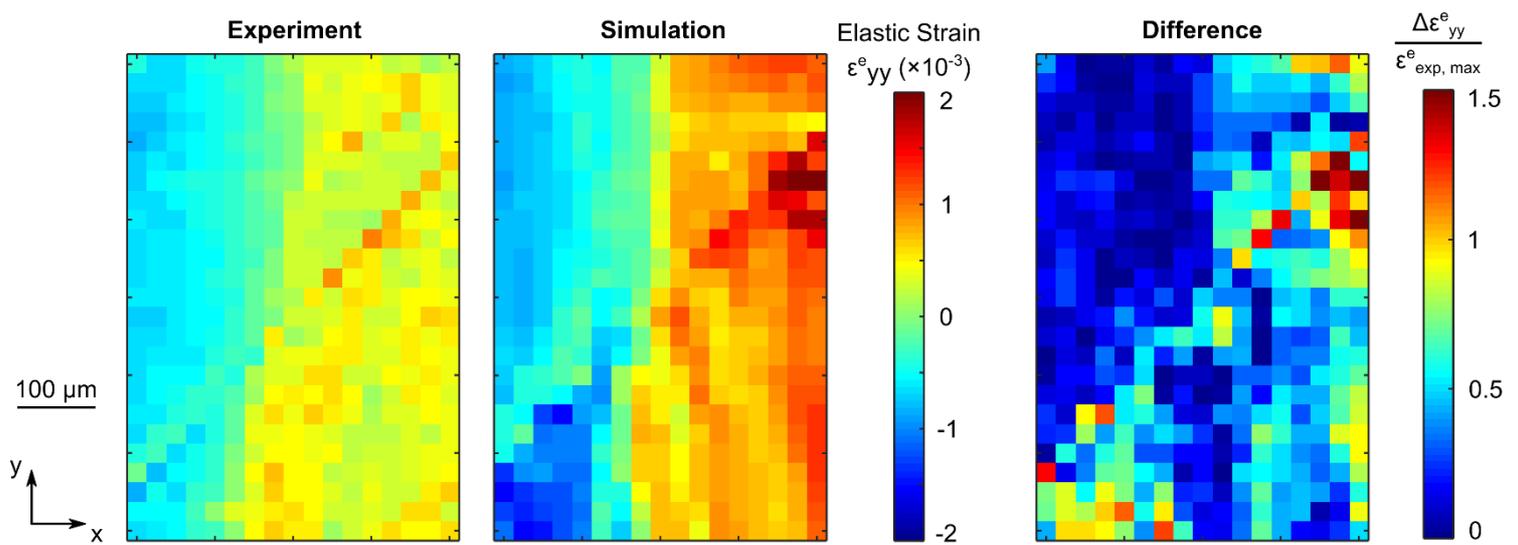

Figure 6: Elastic strain fields along the y axis: experimental, modelling and difference plot at a loading of 7N